\documentclass[preprint,seceq]{jpsj2}
\usepackage{txfonts}
\usepackage{ulem}

\newcommand{\ket}[1]{\left| #1  \right\rangle}
\newcommand{\ave}[1]{\left\langle  #1  \right\rangle}

\newcommand{\CeRhB}{CeRh$_3$B$_2$\ }

\def\mub{\mu_\mathrm{B}}
\title{
Variational Mote Carlo Study of Flat Band Ferromagnetism 
-- Application to CeRh$_3$B$_2$
}

\author{
Hiroshi N. \textsc{Kono}
and Yoshio \textsc{Kuramoto} \thanks{E-mail address: kuramoto@cmpt.phys.tohoku.ac.jp} 
}

\inst{
Department of Physics, Tohoku University, Sendai 980-8578
}

\recdate{\today}

\abst{
A new mechanism for ferromagnetism in CeRh$_3$B$_2$ is proposed
on the basis of variational Monte Carlo results.
In a one-dimensional Anderson lattice where
each $4f$ electron hybridizes with a ligand orbital between neighboring Ce sites,
ferromagnetism is stabilized due to
a nearly flat band which is a mixture of conduction and $4f$ electron states.
Because of the strong spin-orbit interaction in $4f$ electron states, and of considerable amount of hybridization in the nearly flat band, the magnetic moments from $4f$ and conduction electrons tend to cancel each other.
The resultant ferromagnetic moment becomes smaller as compared with the local $4f$ moment, and the Fermi surface in the ferromagnetic ground state is 
hardly affected by the presence of $4f$ electrons.
These theoretical results are consistent with experimental observations in CeRh$_3$B$_2$ by neutron scattering and dHvA effects.
}

\kword{
flat-band ferromagnetism, CeRh$_3$B$_2$, Anderson lattice, variational Monte Carlo
}

\begin{document}
\sloppy
\maketitle

\section{Introduction}

CeRh$_3$B$_2$ has the highest Curie temperature $T_C=120$ K
among known Ce compounds\cite{Dhar_1981,Malik_1982,Yang_1984}. 
If one uses the de Gennes scaling from the transition temperature of 90 K in 
GdRh$_3$B$_2$\cite{Malik_1983}, 
the transition temperature is estimated as small as 1K.
Thus the ordinary RKKY interaction cannot be responsible for the ferromagnetism in CeRh$_3$B$_2$. 
The magnetic moment per formula unit of \CeRhB is 
0.45$\mu_{\mathrm{B}}$\cite{Galatanu_2002}, 
which lies in the $c$ plane.
It is much reduced from the free Ce$^{3+}$ ion value 2.14 $\mu_{\mathrm{B}}$.
Because of these anomalous features the ferromagnetism in CeRh$_3$B$_2$ 
has been attracting much interest.

\CeRhB has the hexagonal CeCo$_3$B$_2$ type structure, whose space group is $P6/mmm (\mathrm{D}_{6h}^1$).
The crystal structure of CeRh$_3$B$_2$  is characterized by
the lattice parameters $a=5.477$ \AA\ and $c=3.091$ \AA.
The lattice constant along the $c$ axis is remarkably short; 
shorter than 3.41 \AA  \ in $\alpha$-Ce with valence close to $4+$.   
Hence a large hybridization between the ligand and $4f$ orbitals, which is called the $c$-$f$ hybridization, and a quasi-one dimensional band structure are expected. 
Previous studies actually suggest the quasi-one-dimensional feature along the $c$-axis\cite{Okubo_2003,Harima_2004,Yamada_2004}. 

Several models have been proposed to explain the anomalous ferromagnetism. 
However, mechanism of ferromagnetism in CeRh$_3$B$_2$ has not yet been identified.
For example, an itinerant model of ferromagnetism due to Rh 4$d$ can explain small magnetic moment\cite{Dhar_1981,Malik_1983}. 
It was considered as support of the model
that no magnetic order is observed in CeRu$_3$B$_2$\cite{Dhar_1981}.
However, the model 
cannot explain the fact that Rh $4d$ bands of LaRh$_3$B$_2$ and CeRh$_3$B$_2$ look essentially the same according to valence-band photoemission 
\cite{Sampathkumaran_1985},
while LaRh$_3$B$_2$ shows Pauli paramagnetism and superconducting transition at about 2.4K\cite{Ku_1980}. 

On the other hand, a $4f$ itinerant model can also explain the small ferromagnetic moment\cite{Malik_1983}. 
It is a likely scenario, since the extreme proximity of Ce ions along the $c$ axis 
may cause some delocalization of $4f$ charge.
Batista \textit{et al.} have in fact proposed that the ferromagnetism of CeRh$_3$B$_2$ can be explained by the periodic Anderson model (PAM)\cite{Batista_2002}. 
In their work, the high Curie temperature and the small magnetic moment are obtained.  
Recently, de Haas-van Alphen (dHvA) effects have been observed for LaRh$_3$B$_2$ and CeRh$_3$B$_2$\cite{Okubo_2003}.
The dHvA results suggest that $4f$ electron is localized, which is in conflict with the $4f$ itinerant ferromagnetism models.
As a result, a localized electron model has also been considered as a likely model. However, the small magnetic moment is difficult to be explained by this model.

In present work, we propose a new model for the ferromagnetism in CeRh$_3$B$_2$ with special attention to the role of {\it c-f}
hybridization and the orbital moment of $4f$ electrons.  
According to the band calculation \cite{Harima_2004}, the relevant conduction band is regarded as quasi one-dimensional, and is mainly formed by molecular orbitals of Rh $4d$ states lying between Ce sites along the $c$-axis.
We therefore take the one-dimensional 
Anderson lattice with hybridization between the nearest $4f$-ligand orbitals. 
To analyze the model, we use the Optimization Variational Monte Carlo(O-VMC) method. 
We determine the ground state phase diagram 
and analyze the mechanism of ferromagnetism in terms of the effective one electron band which is obtained by O-VMC.
On the basis of these results, we 
discuss the mechanism of the ferromagnetism of CeRh$_3$B$_2$. 
In this paper we consider properties of the ground state only.

\section{Model}

The spin-orbit interaction leads to the total angular momentum $J=5/2$ for the lowest $4f^1$ state. 
The six fold degeneracy is split into three 
Kramers doublets by the crystal field.
These states are represented in terms of the basis $\ket{J_z}$ of $J=5/2$ as follows:
 \begin{align}
\ket{\pm 5/2} &=\pm \sqrt{\frac{6}{7}} \ket{\pm3,\mp1/2} \mp \sqrt{\frac{1}{7}} \ket{\pm2,\pm1/2}, \\
\ket{\pm 3/2} &=\pm \sqrt{\frac{5}{7}} \ket{\pm2,\mp1/2} \mp \sqrt{\frac{2}{7}} \ket{\pm1,\pm1/2}, \\
\ket{\pm 1/2} &=\pm \sqrt{\frac{4}{7}} \ket{\pm1,\mp1/2} \mp \sqrt{\frac{3}{7}} \ket{0,\pm1/2}\label{updown},
\end{align}
where $\ket{L_z, S_z}$ stands for a basis of $L=3$ and $S=1/2$.

The crystal-field ground state is taken as $\ket{\pm 1/2}$\cite{Yamada_2004}. 
We keep only the doublet $\ket{\pm 1/2}$ for explicit calculation including hybridization.
This simplification is motivated by
the high excitation energy 
of 220K between $\ket{\pm 1/2}$ and $\ket{\pm 3/2}$\cite{Yamada_2004}, 
which is about two times the transition temperature.
When the $4f^1$ states are restricted to $\ket{\pm1/2}$, the 
spin-orbit interaction makes
the $xy$ plane an easy plane, and the $z$ axis the hard axis.
To clarify this point, we consider the wave functions 
$\Psi_{x\sigma}$ where the moment is along the $x$ axis.
We obtain
\begin{align}
\Psi_{x\uparrow} =&\frac{1}{\sqrt{2}}(\ket{1/2}+\ket{-1/2}).
\end{align}
The expectation values of moments are obtained as
\begin{equation}
\langle\Psi_{x\uparrow}|J_x|\Psi_{x\uparrow}\rangle =3/2, \ 
\langle 1/2|J_z|1/2\rangle =1/2.  
\end{equation}
Hence the easy axis lies in the $xy$ plane.

\begin{figure}
 \begin{center}
\includegraphics[width=8cm,keepaspectratio]{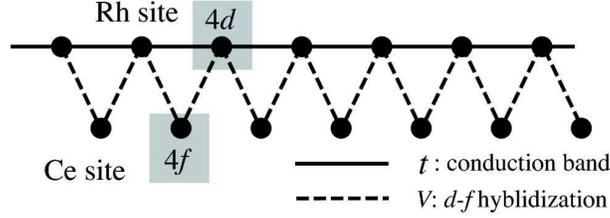}
 \caption{
Illustration of the model in which each $f$ orbital hybridizes with nearest Rh 4$d$ molecular orbitals.   
 }
 \label{modelI}
 \end{center}
\end{figure}

According to dHvA experiment \cite{Okubo_2003} and the band structure calculation \cite{Harima_2004}, the Fermi surface of LaRh$_3$B$_2$ has the quasi-one-dimensional feature and 
consists of Rh $4d$ orbitals.
It has been  pointed out that the origin of the quasi-one-dimensional feature is the strong hybridization between Rh $4d$ bands along the $c$ axis \cite{Okubo_2003,Harima_2004}.
We adopt the one dimensional model where 
the lowest CEF orbital of $4f$ states hybridizes with
a molecular orbital formed by six Rh sites surrounding 
the $c$ axis threading Ce ions.
The center of the molecular orbital sits at the mid point between the Ce sites.
We take into account the nearest neighbor hybridization between Rh $4d$ and Ce $4f$ states.
Figure \ref{modelI} illustrates our model.

For each site we 
introduce a creation operator $d_{\sigma}^\dagger$ for
the Rh $4d$ molecular orbital 
and $ f_{\sigma}^\dagger$ for the Ce CEF states,
where $\sigma =\uparrow, \downarrow$ specifies
the the eigenstates of the spin operator $S_x$.
We note that 
$\langle\mib J\rangle$ is 
anti-parallel to $\langle\mib S\rangle$ in the Hund-rule ground state.
Hence the spin points to the opposite direction of the total moment $J_x$ of the $4f$ electron.
Our model is written as follows:
\begin{align}
\mathcal{H}=& \sum_{\langle i j \rangle \sigma} t d_{i\sigma}^\dagger d_{j \sigma} +\sum_{i \sigma} \epsilon_d d_{i \sigma}^\dagger d_{i\sigma}+\sum_{i \sigma} \epsilon_f f_{i\sigma}^\dagger f_{i\sigma} \nonumber \\
&+\sum_{[i j] \sigma} V ( f_{i\sigma}^\dagger d_{j \sigma} + d_{j\sigma}^\dagger f_{i\sigma} )
+\sum_{i} U n_{i\uparrow}^f n_{i\downarrow}^f,
\label{4d-4f}
\end{align}
where $\langle i j \rangle$ denotes a nearest neighbor Rh orbital pair
with hopping energy $t$,
 and $\sum_{[i j]}$ is the summation 
 over nearest $4f$  and $d$ sites with 
hybridization matrix element $V$.
We note that the distance between the nearest $f$ and $d$ sites is half of the lattice spacing.
The energy of the the $4f$ orbital is written as
$\epsilon_f$,  and that of the $d$ orbital as
$\epsilon_d$, which is taken to be 0.
Namely we take the origin of the energy 
at the center of the conduction band. 
We take $t=1$ as the unit of energy.  
The positive value puts the band bottom at the edge of the Brillouin zone, 
which is consistent with  the band structure of LaRh$_3$B$_2$.

\begin{figure}
 \begin{center}
\includegraphics[width=8cm,keepaspectratio]{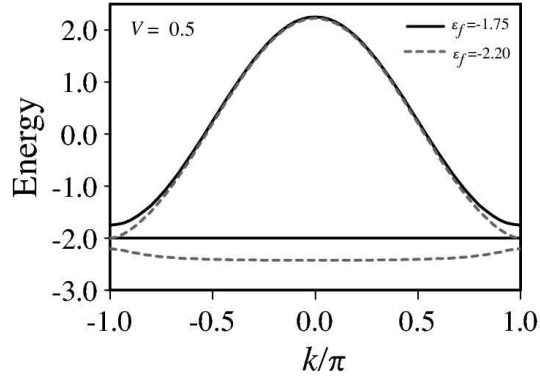}
 \caption{ 
Examples of the energy band structure. The lower band becomes completely flat for $\epsilon_f= V^2/t -2t$.
 }
 \label{flatband}
 \end{center}
\end{figure}

This model has a characteristic 
band structure in the noninteracting limit $U=0$.
Namely the lower 
hybridized band becomes completely flat 
with $\epsilon_f= V^2/t -2t$.
We also note that the hybridization term becomes zero 
for $k=\pi$, because 
of the form $2V\cos(k/2)$ in the momentum space.
Hence eigenvalue of Eq. (\ref{4d-4f})  becomes $-2t$ and $\epsilon_f$. 
Figure \ref{flatband} shows the flat band structure together with a case of slightly shifted $\epsilon_f$ 
for $t=1$ and $V=0.5$.

\section{Variational Monte Carlo Method}

It has been proven in the case of equal number of electrons and unit cells
that 
the ground state of this kind of models   
is ferromagnetic, provided the flat band condition is satisfied 
\cite{Mielke_1993_2} or nearly so \cite{Tasaki1993}. 
Ferromagnetism in two-band models near the flat band condition have already been discussed in great detail 
by a variety of methods \cite{Penc_1996, Guerrero_2001}.
In the preset work, we use a variational Monte Carlo (VMC) to investigate the ferromagnetism in CeRh$_3$B$_2$.
In contrast with previous VMC work \cite{Yokoyama_1997} for two-band ferromagnetism, however, our O-VMC takes account of spin dependence of effective hybridization and local level.
As a result, our O-VMC has an advantage to provide
the effective band picture, which allows an intuitive explanation how the 
particular value of polarization is stabilized.
This picture is especially useful for the case where the flat band condition is not satisfied strictly.
Hence it is expected that the present work should bring new insight into the ferromagnetism near the flat band condition.

We use 
an optimization technique\cite{Umrigar_1998} in the variational Monte Carlo (O-VMC) method.
This technique allows us to introduce five variational parameters.
To consider a ferromagnetic state, we introduce spin dependent effective hybridization parameter $\tilde{V}_\sigma$ and effective $4f$ level $\tilde{\epsilon}_{f \sigma}$.
Using these parameters, we 
construct the variational state as follows:
We start from Eq. (\ref{4d-4f}) with $U=0$ . 
By 
replacing $V$ and $\epsilon_f$ by $\tilde{V}_\sigma$ and $\tilde{\epsilon}_{f \sigma}$ respectively, we 
obtain
\begin{align}
\tilde{\mathcal{H}}_0=& \sum_{\ave{i j} \sigma} t d_{i\sigma}^\dagger d_{j \sigma} +\sum_{i  \sigma} {\epsilon}_{d, \sigma}d_{i \sigma}^\dagger d_{i\sigma}\nonumber \\
&+\sum_{i  \sigma} \tilde{\epsilon}_{f \sigma} f_{i\sigma}^\dagger f_{i\sigma} +\sum_{[i j] \sigma} \tilde{V}_\sigma ( f_{i\sigma}^\dagger d_{j \sigma} + d_{j\sigma}^\dagger f_{i\sigma} ).
\label{v_model}
\end{align}
We diagonalize $\tilde{\mathcal{H}}_0$
and derive the upper 
and lower hybridized bands 
which depend on the variational parameters as
\begin{align}
E_{k \sigma \pm} =\frac{1}{2} \left(\epsilon_k +\tilde{\epsilon}_{f \sigma} \pm \sqrt{(\epsilon_k -\tilde{\epsilon}_{f\sigma})^2 +16\tilde{V}_{\sigma}^2 \cos^2 {k}/2} \right),
\end{align}
where $\epsilon_k = 2t \cos{k}$.
The creation operator of the upper (lower) hybridized band state 
$a_{k \sigma}^\dagger$ ($b_{k \sigma }^\dagger$) is written as 
\begin{align}
a_{k \sigma }^\dagger &= u_{k\sigma} d_{k\sigma}^\dagger + v_{k\sigma} f_{k \sigma}^\dagger,\\
b_{k \sigma }^\dagger &= -v_{k\sigma} d_{k\sigma}^\dagger + u_{k\sigma} f_{k \sigma}^\dagger,
\label{generation}
\end{align}
where $u_{k\sigma}^2$ and $v_{k\sigma}^2$ are 
weights of the $4f$ and $d$ states 
in the lower band: 
\begin{align}
u_{k\sigma} &= \frac{\epsilon_k - E_{k\sigma  -}}{\sqrt{4V^2\cos^2 {k}/2+(\epsilon_k-E_{k\sigma -})^2}},\\
v_{k\sigma} &= \frac{2V\cos {k}/2}{\sqrt{4V^2\cos^2 {k}/2+(\epsilon_k-E_{k\sigma -})^2}}.
\end{align}

For a ferromagnetic state, we take
the number $N_{\uparrow}$ of up spins larger than
the number $N_{\downarrow}$ of down spins without loss of generality.
The total electron number $N_e$ is given by
$N_e  = N_{\uparrow} +N_{\downarrow}$.
We first prepare an noninteracting state 
to construct a variational wave function.
With $N_{\uparrow} \leq L$, the magnetic state 
is given by
\begin{align}
\ket{\Phi} = \prod_\sigma \prod _ {{k}}^{N_{\sigma}} b_{k \sigma }^\dagger \ket{0}
\label{no_int_m}
\end{align}
where $\ket{0}$ is the vacant state,
and $N_\sigma$ over the product symbol means the number of $k$'s involved. 
On the other hand, with $N_{\sigma} > L$ 
for both spins, electrons occupy not only the lower band  but also the upper band.
Hence the noninteracting magnetic state 
is given by
\begin{align}
\ket{\Phi} = 
\prod_\sigma  
\prod _ {{k}}^{N_{\sigma}-L} a_{k \sigma }^\dagger 
\prod _ {{k}}^{L} b_{k \sigma }^\dagger 
\ket{0},
\label{no_int_m}
\end{align}
We have also considered a more general case where the upper hybridized band of up spins is partially occupied even though
the lower hybridized band of down spins is not full.
Starting from one of these noninteracting states, we construct variational wave functions by operating the Gutzwiller projection: 
\begin{align}
P= \prod_{i} (1-\tilde{\eta} n^f_{i \uparrow} n^f_{i \downarrow}),
\end{align}
where $\tilde{\eta}$ is a variational parameter 
to restrict double occupancy.
If $\tilde{\eta} =1$, 
any state with double occupation is excluded, which
is consistent with $U= \infty$.
On the other hand, if $\tilde{\eta} =0$, the 
Gutzwiller projection operator becomes the identity operator $P=1$, which
corresponds to the case of  $U=0$.
The variational wave function $\ket{\Psi}$
with the Gutzwiller projection  
is given by
\begin{align}
\ket{\Psi} = P \ket{\Phi},
\label{GWF}
\end{align}
where the variational parameters are included in both $\ket{\Phi}$ and $P$.

\section{Numerical Results for Ferromagnetism}

\begin{figure}
 \begin{center}
\includegraphics[width=8cm,keepaspectratio]{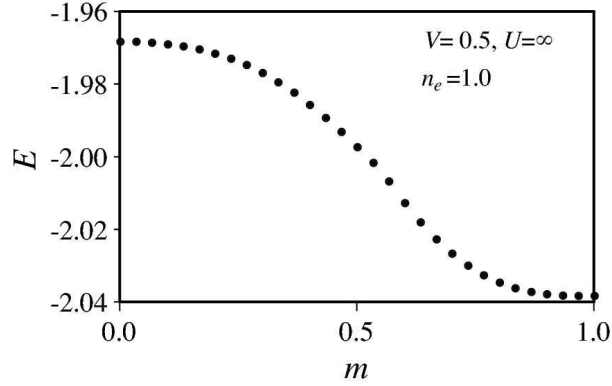}
 \caption{Energy 
 as a function of magnetization in the case of $L=120$.
 }
 \label{ne1Uinfef-18EnvsM}
 \end{center}
\end{figure}

We define the magnetization $m$ per site as 
$m=(N_{\uparrow}-N_{\downarrow})/L$, and the density as
$n_e = (N_{\uparrow}+N_{\downarrow})/L$, 
where $L$ is the total number of unit cells.
We take $V=0.5$ and either $\epsilon_f=-1.8$ or $\epsilon_f=-2.2$ to determine  the phase diagram in the plane of $n_e$ and $U$.
Figure \ref{ne1Uinfef-18EnvsM} shows the energy per unit cell
as a function of magnetization with $n_e=1$ and $\epsilon_f=-1.8$.
We have taken the system with size $L=120$ and the  anti-periodic boundary condition.  
The minimum of the energy 
is located at $m=n_e$.
In this case, the ground state is a ferromagnetic insulator and 
is fully polarized. 
We have done similar calculation taking $L=40$
and found almost identical result with that in Fig.\ref{ne1Uinfef-18EnvsM}.  
Therefore in the following we present results with $L=40$ for various choices of $n_e$ and $U$.

We classify the ferromagnetic ground state into three kinds:\\ 
(i) The fully  polarized state where all spins align.  This state appears at  $n_e=1$.\\
(ii) A band ferrimagnetic state which appears with $1<n_e<2$.  Here
the lowest effective band is fully polarized, and the next band is polarized in the opposite direction.  Hence we have 
$m=1-(n_e-1)=2-n_e$.   This state is also called the complete ferromagnetism in the literature \cite{Guerrero_1996}.\\
(iii) A ferromagnetic state where the lowest band is fully polarized and  the effective higher bands have no polarization.
The magnetization of this state is $m=1$, and appears
in the region with high $U$ and low electron density of the phase diagram for $\epsilon_f=-2.2$. \\
(iv) Other ferromagnetic states which generally have $m>2-n_e$.   In this case the lowest band is fully polarized, but other bands are partially filled.  

\begin{figure}[bp]
 \begin{center}
\includegraphics[width=8cm,keepaspectratio]{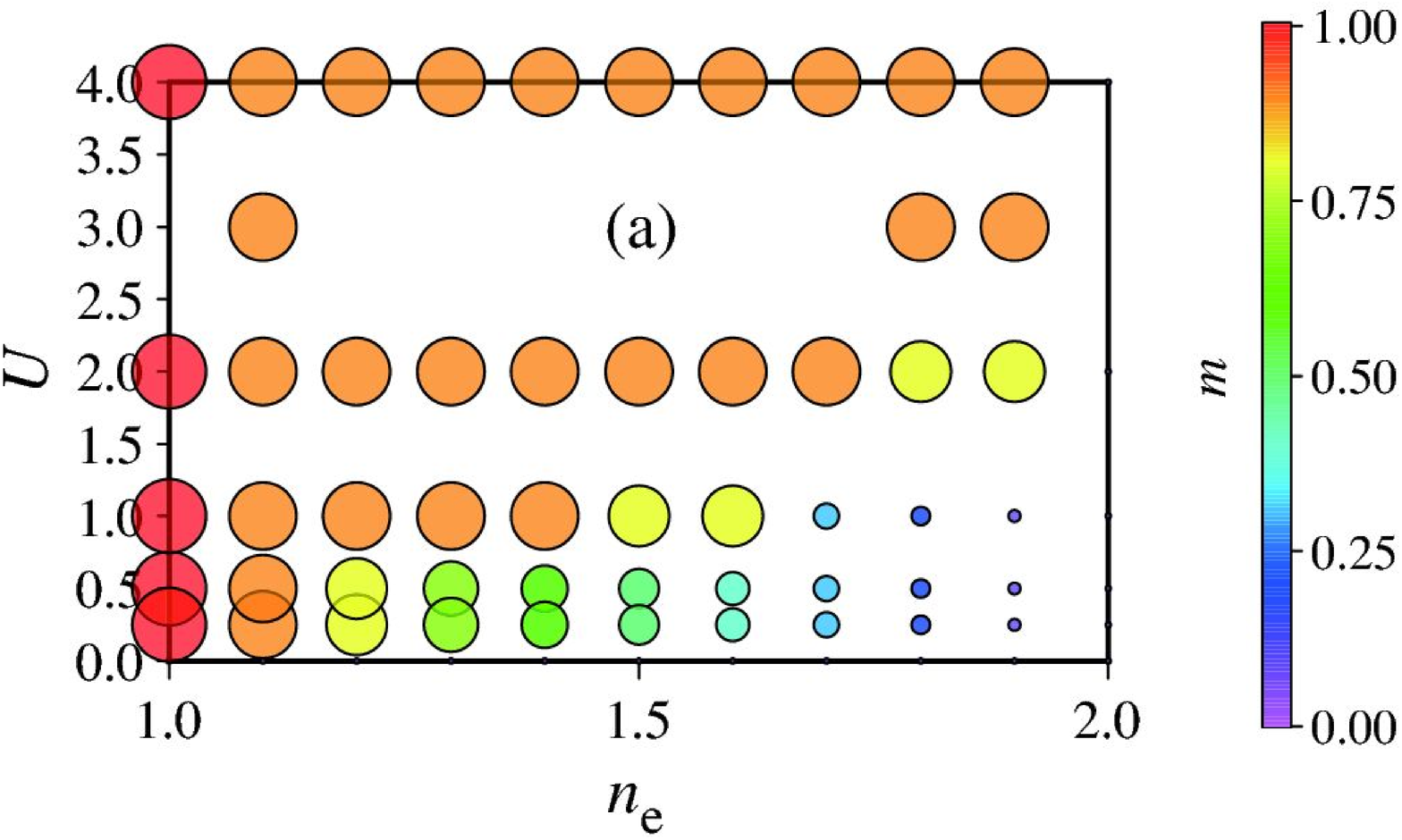}
\includegraphics[width=8cm,keepaspectratio]{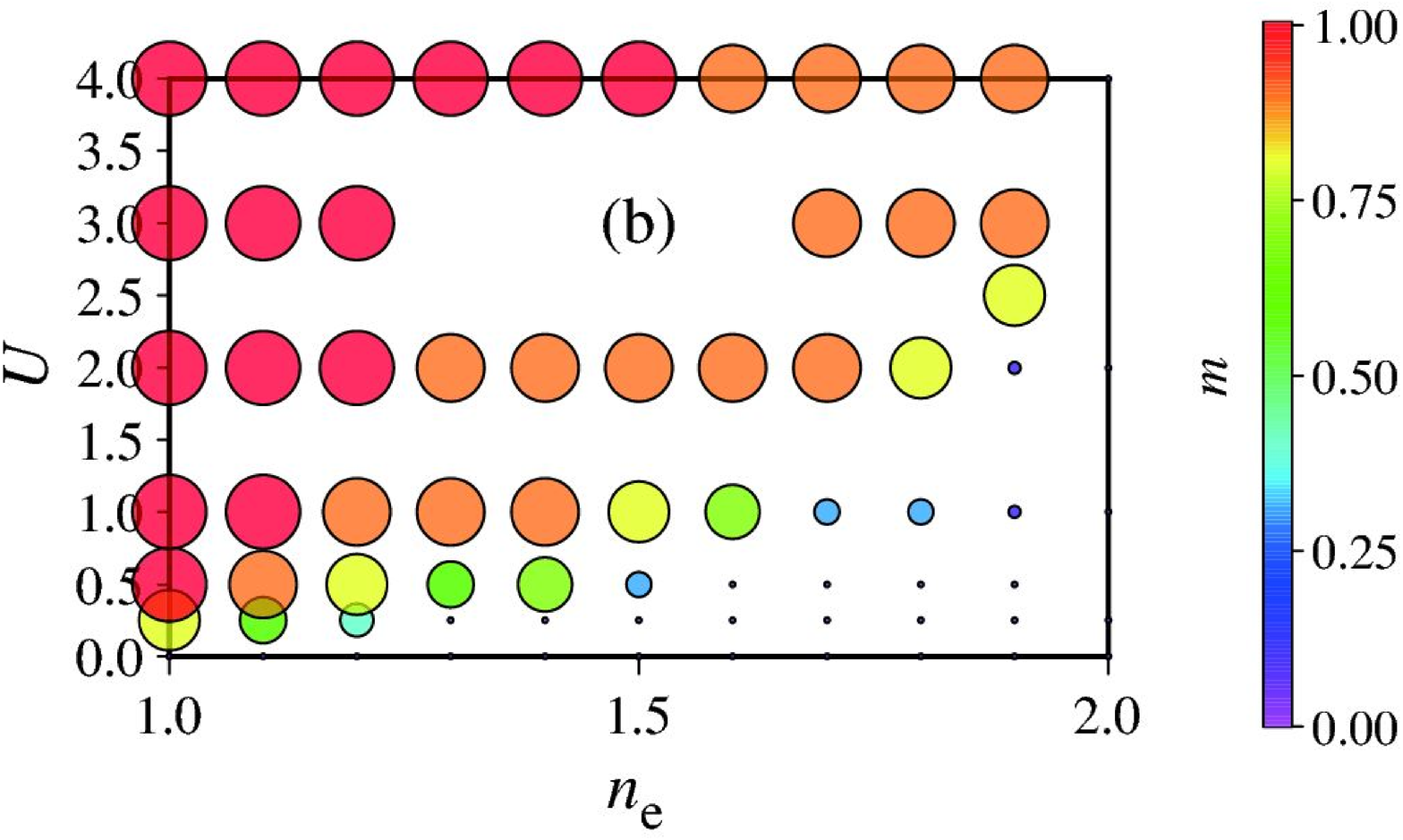}
\caption{
 Ground state phase diagram in the plane of 
$n_e$ and $U$ for
 (a) $\epsilon_f=-1.8$, and 
 (b) $\epsilon_f=-2.2$.
The radius of each circle represents the magnitude of magnetization.
The dashed line in (a) separates the regions (ii) and (iii) as explained in the text.}
 \label{VMC_Phase_22}
 \end{center}
\end{figure}

Figure 
\ref{VMC_Phase_22} 
shows phase diagram with $L=40$
for 
(a) $\epsilon_f=-1.8$ and 
(b) $\epsilon_f=-2.2$.
It turns out that the ground state can be ferromagnetic even in the region away from $n_e=1$. 
For $\epsilon_f=-2.2$, the region with $m=1$ is larger than that for $\epsilon_f=-1.8$.
Additionally the paramagnetic region expands for smaller $U$.

\begin{figure}
 \begin{center}
\includegraphics[width=8cm,keepaspectratio]{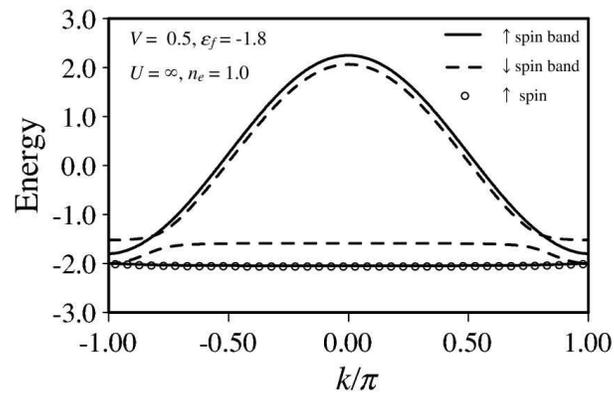}
 \caption{The effective band for $n_e=1.0$ and $U=\infty$. The circles 
 represent occupied states of up spin electrons.}
 \label{ne10effectiveband}
 \end{center}
\end{figure}

To give a physical explanation of the ferromagnetism, we 
derive the effective band structure in the ferromagnetic state.
First we consider the case of 
$n_e=1.0$ and $U=\infty$, which is given in Fig. \ref{ne10effectiveband}.
The optimized variational parameters are given as $\tilde{V}_{\uparrow}=0.5$, $\tilde{\epsilon}_{f\uparrow}=-1.8$, $\tilde{V}_{\downarrow}=0.243$ 
and $\tilde{\epsilon}_{f\downarrow}=-1.52$.
The lower up-spin band is not 
affected by the Coulomb interaction,
while the lower down-spin band is 
uplifted.
As a result
all electrons occupy the lower up-spin band.

\begin{figure}
 \begin{center}
\includegraphics[width=8cm,keepaspectratio]{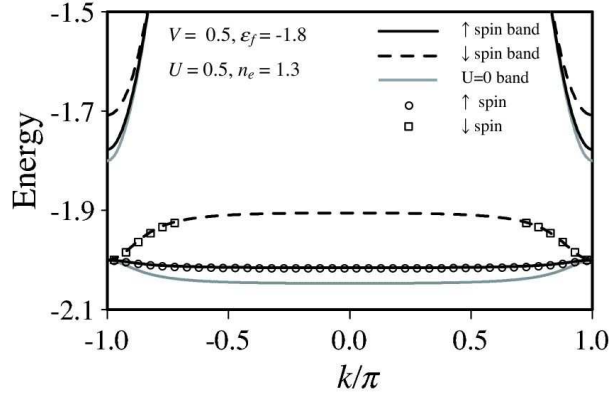}
 \caption{The effective band for $n_e=1.3$ and $U=0.5$. 
 The circles (squares)  represent the occupied states of up (down) spin electrons.}
 \label{ne13effectiveband}
 \end{center}
\end{figure}

Next we derive
the effective band for $n_e=1.3$ and $U=0.5$, which is given in Fig. \ref{ne13effectiveband}.
The lower down spin-band is affected by the Coulomb interaction and is uplifted as in $n_e=1.0$. 
Since the electron number is larger than the full occupancy of the lowest band,
the extra electrons occupy the lower down-spin band.
Thus 
the total magnetic moment has the magnitude $m=2-n_e$.
Namely 
the band ferrimagnetic state is realized.
Since 
the parameters are close to  the flat band condition,
one may interpret this ferromagnetism as 
a generalized flat band ferromagnetism.

\begin{figure}
 \begin{center}
\includegraphics[width=8cm,keepaspectratio]{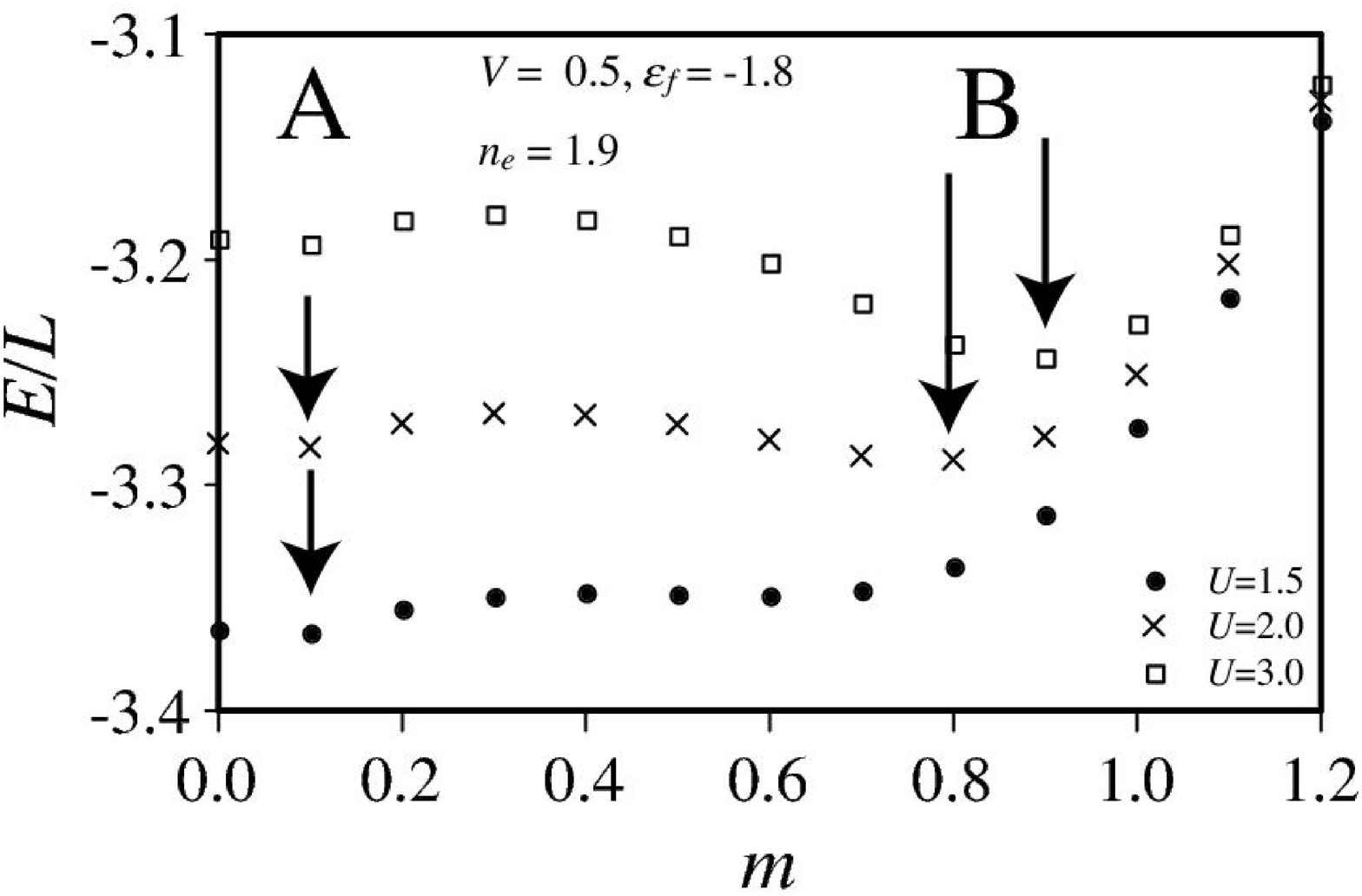}
 \caption{Energy 
  as a function of magnetization for $U=$1.5, 2, and 3 with $L=40$.
 }
 \label{ne11effectiveband}
 \end{center}
\end{figure}

On the boundary between 
(ii) and (iii) of the ground state phase diagram,
the magnetization changes discontinuously. 
To clarify the behavior near the boundary, we compute the energy 
as a function  of the magnetization for $U=$1.5, 2.0, and 3.0.
This result is shown in  Fig. \ref{ne11effectiveband} .
For $U=1.5$, the energy 
has the minimum at $m=0.1$ (Point A). 
As $U$ becomes larger, 
another local minimum appears in the vicinity of $m=0.8$ (Point B) 
The point B becomes 
the absolute minimum for $U\geq2.0$.
Hence the magnetization changes  discontinuously at $U\sim 2.0$. 

\begin{figure}
 \begin{center}
\includegraphics[width=8cm,keepaspectratio]{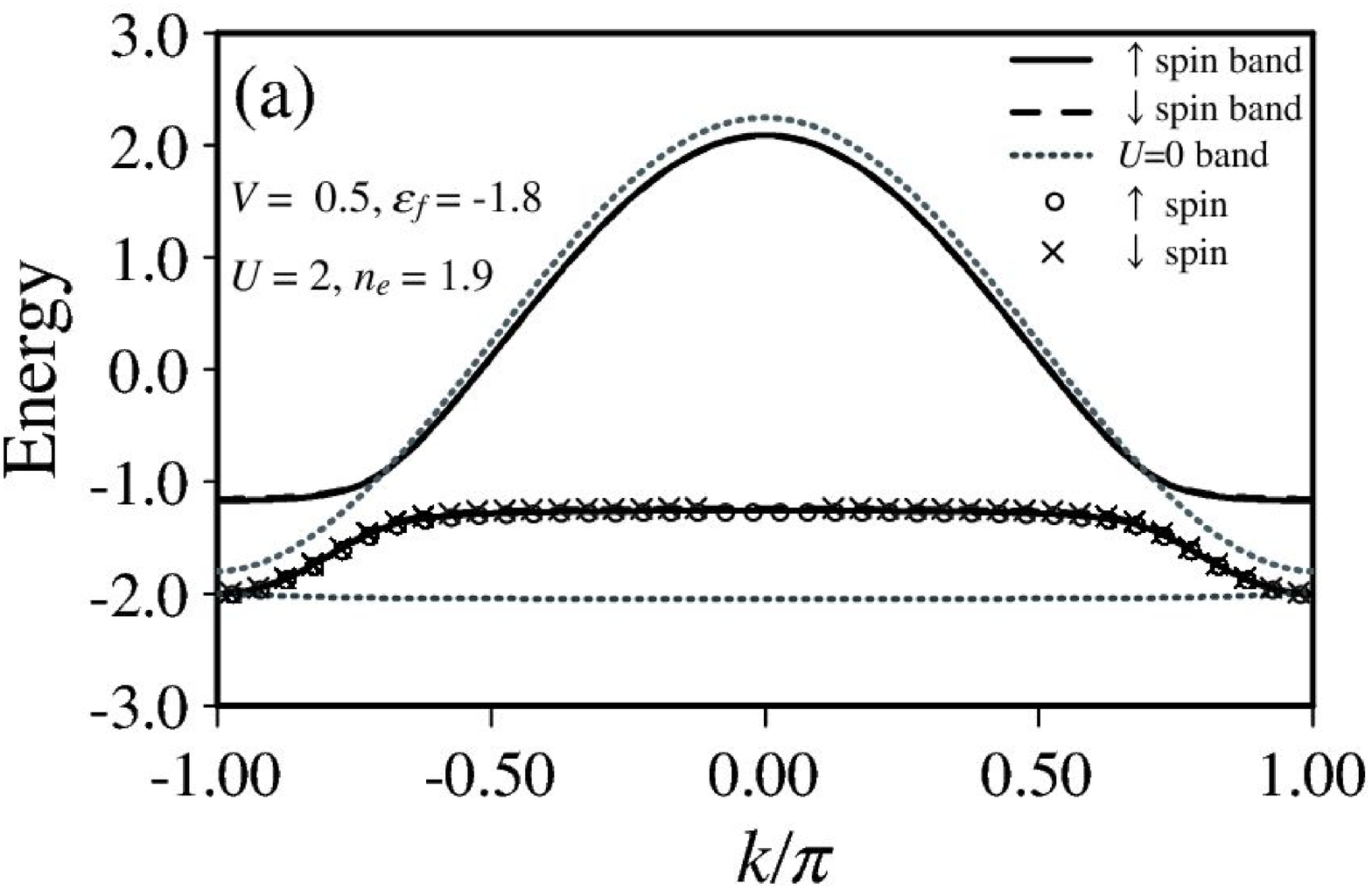}
\includegraphics[width=8cm,keepaspectratio]{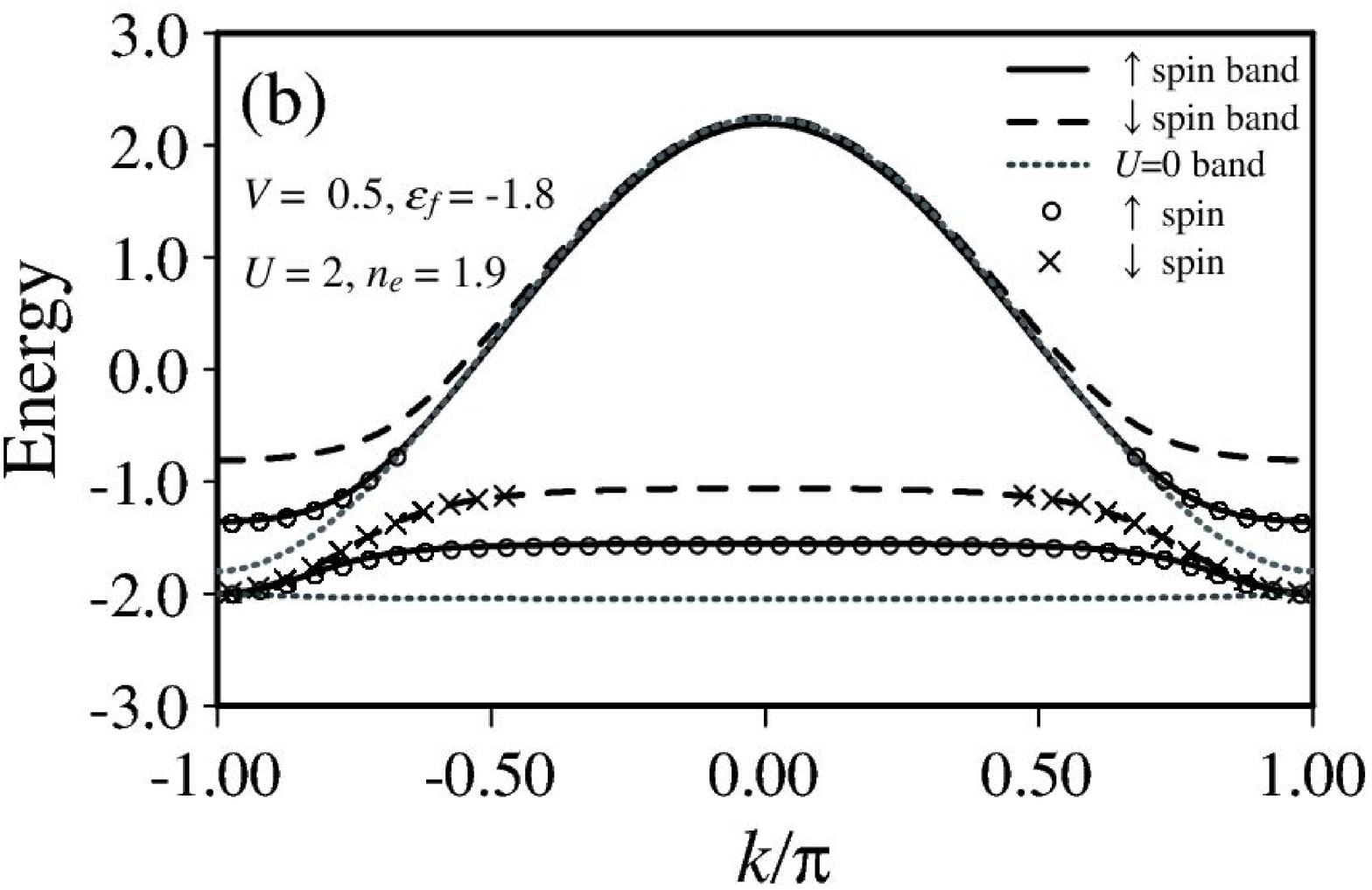}
 \caption{The effective bands for $n_e=1.9$ and $U=2$ with
(a) $m=0.1$ corresponding to the point A in Fig.\ref{ne19effectiveband_B}, and
(b) $m=0.8$ corresponding to the point B.}
 \label{ne19effectiveband_B}
 \end{center}
\end{figure}

We also derive the effective band structure at point A and B for $U=2.0$.
For $U=2.0$, point A and B have almost the same energy. 
These results are shown in  
Fig. \ref{ne19effectiveband_B}.
At point  A, the ground state is ferromagnetic which is connected to the flat band ferromagnetism.
The effective lower band shifts to the higher energy compared to the one for $U=0$.
In the center of the lower energy band, the $4f$ component is dominant and there 
appears almost localized states.
These states become unstable as $U$ becomes larger because 
the double occupation is not excluded completely.

On the other hand, at point B, the electrons occupy not only the 
lower hybridized bands but also the upper band of up spins.
The effective bands which consist mainly of $4f$ orbital splits into up and down spin bands so as to reduce the double occupation.
Thus the discontinuity of the magnetization is caused by the transition between the flat band ferromagnetism at point A and the 
almost localized ferromagnetism at point B.
However, in reality some antiferromagnetism may also appear
 in the large $U$ region.
Since our variational wave function does not  involve the antiferromagnetic correlation adequately, we cannot determine the most stable magnetic state by our approach.

\begin{figure}
 \begin{center}
\includegraphics[width=8cm,keepaspectratio]{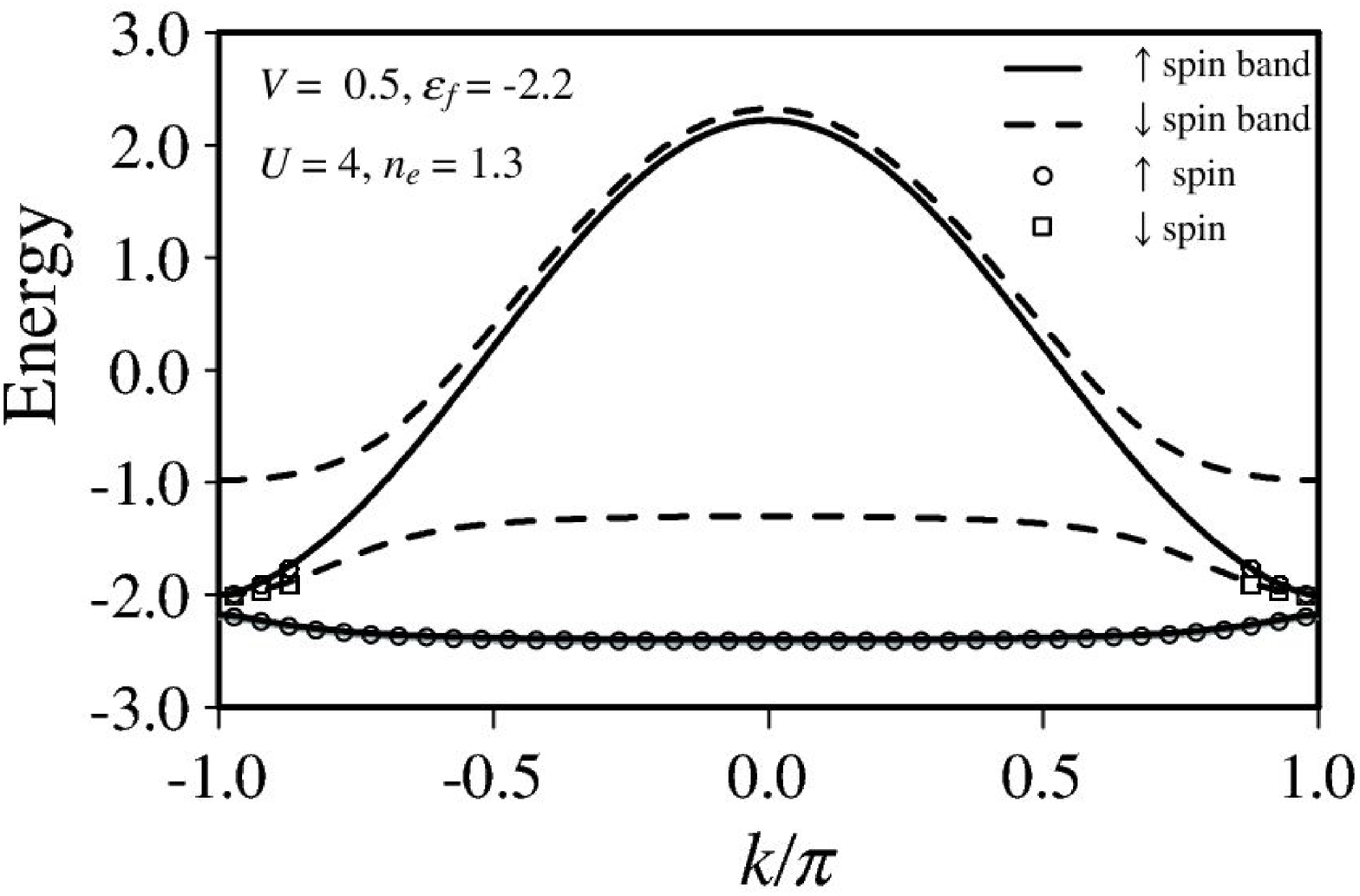}
 \caption{The effective band for $\epsilon_f = -2.2$, $n_e=1.3$ and $U=4$. 
 The circles (squares)  represent the occupied states of up (down) spin electrons.}
 \label{ne13effectiveband22}
 \end{center}
\end{figure}

We  have also derived the effective band for $\epsilon_f=-2.2$, $n_e=1.3$ and $U=4$, which is shown in Fig. \ref{ne13effectiveband22}.
The Fermi level is located in the effective upper band.
We find the optimum state such that the lower hybridized band of 
down spins and the upper hybridized up spins have the same Fermi wave number. 
Namely, there is no polarization at the Fermi surface, and the value of $k_F$ agrees with those of the localized $4f$ electron model.
In contrast with the localized model, however, the fully polarized band
involves the $4d$ component.
Since we have taken the system size $L=40$ in the calculation, we cannot exclude the possibility that there is a very weak polarization at the Fermi surface.

\section{Application to CeRh$_3$B$_2$}

We compare the ferromagnetism of this model with experimental 
results for CeRh$_3$B$_2$.
For this purpose we use the following parameters:
 $t$=0.34eV, $U$=7eV, $\epsilon_f=-0.714$eV, $V$=0.24eV and $n_e=1.1$.
Here $t$ is determined by comparison with the band structure calculation\cite{Harima_2004}.
It is difficult to estimate the hybridization between the molecular orbital and the $4f$ orbital.
On the other hand,  hybridization averaged over the band 
is estimated by X-ray absorption spectroscopy (XAS)
to be between 0.23 and 0.4 eV \cite{Jo_1990,Fujimori_1990,Yamaguchi_1995}. 
We tentatively adopt the value 0.24eV for $V$.
The value $n_e=1.1$ is determined by the volume of the Fermi surface which is derived by the band structure calculation \cite{Harima_2004}.
Then we find that the ground state shows the complete polarization in the lowest band, and almost no polarization in other bands.  
The important features are as follows::

\noindent (i) Anisotropy of the magnetization\\
The strong anisotropy of the magnetic moment can be explained by taking the crystal field state as $J_z=\pm 1/2$.  In reality, other components of $J_z$ may be contributing to the ground $4f^1$ level,
and may influence the anisotropy and the magnitude of magnetization.

\noindent (ii) Polarization of $4d$ electrons\\
In the region close to the flat band condition, the lower band 
polarizes completely.  Therefore $4d$ electrons also polarize significantly with strong hybridization. 
Because of the spin-orbit interaction of $4f$ electrons, 
the magnetic moment of $4d$ electrons is anti-parallel to that of $4f$ electrons.
The magnetic moment per $4d$ electron is derived as $0.15\mub$ by our calculation, which is close to the value $0.18\mub$ estimated by neutron scattering experiment. 

\noindent (iii) Reduced moment at Ce sites \\
The total magnetic moment is estimated as 0.94$\mub$, 
which is larger than the experimental result 0.45$\mub$.
A possible source of difference is that the crystal field state assumed in the present work does not represent the actual electronic state. 
In order to reproduce the observed value of the moment, the expectation value of $J_x$ should be about 0.6 instead of 3/2 in the present model.

\begin{figure}
 \begin{center}
\includegraphics[width=8cm,keepaspectratio]{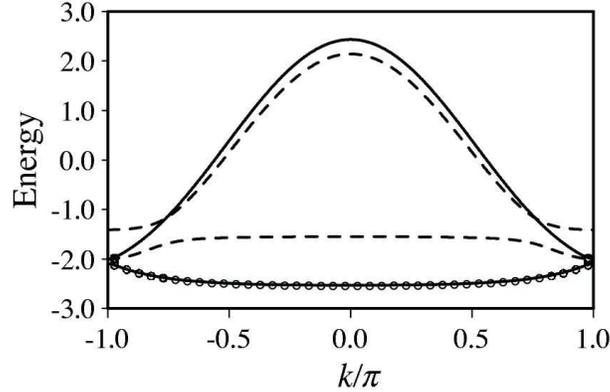}
 \caption{The effective band for $t$=0.34eV, $U$=7eV, $\epsilon_f=-0.714$eV, $V=0.24$eV and $n_e=1.1$.}
 \label{exp}
 \end{center}
\end{figure}

\noindent (iv) Fermi surface \\
The effective band is shown in Fig. \ref{exp}
where the Fermi level is located near the bottom of the effective upper band with the Fermi wave numbers $k_{F\sigma}/\pi=\pm 0.975$.
This situation 
corresponds to the case shown in Fig. \ref{ne13effectiveband22}.
There is no polarization at the Fermi surface, and the value of $k_F$ agrees with those of the localized $4f$ electron model.

\noindent (v) Curie temperature\\
Although our approach cannot discuss the Curie temperature directly, we can estimate the Curie temperature by
the difference between the paramagnetic state ($m=0$) and the  
ferromagnetic state.
With the choice of the optimum magnetization
$m=1$, the difference is calculated to be 826 K per site with the parameters used.
This is surely the overestimate since our variational wave function in the paramagnetic state does not take proper account of correlation effects such as antiferromagnetic fluctuations.   However, our result does show that the present mechanism of the ferromagnetism is consistent with the high Curie temperature of CeRh$_3$B$_2$.

\section{Conclusion}

In present work, we 
have applied the O-VMC to study 
the mechanism of ferromagnetism near the flat band condition. 
In the region both at and away from $n_e=1$,  
we have found that the band ferromagnetism is stabilized.
We have compared the ferromagnetism of this model with the experimental results for CeRh$_3$B$_2$.
We have provided reasonable explanation for the anisotropic magnetization,
reverse polarization of conduction electrons, and the size of the Fermi surface probed by the dHvA effect.
On the other hand, the total moment obtained by our model is much larger than the experimentally reported value.  A possible source of difference is ascribed to our CEF state which may not be justified quantitatively.
It is hoped that improvement of the present model will provide more quantitative explanation of the ferromagnetism in CeRh$_3$B$_2$.

\section*{Acknowledgments}
We would like to thank H. Yokoyama, N. Sibata and H. Kusunose for valuable discussion.
The authors thank the Supercomputer Center, Institute for Solid State Physics, University of Tokyo for the use of the facilities.


\begin{thebibliography}{99}
\bibitem{Dhar_1981} S.K. Dhar, S.K. Malik and R. Vijayaraghavan: J. Phys. C {\bf 14} (1981) L321.
\bibitem{Malik_1982} S.K. Malik, S.K. Dhar, R. Vijayaraghavan and W.E. Wallace, J. Appl. Phys. {\bf 53} (1982) 8074.
\bibitem{Yang_1984} K.N. Yang, M.S. Torikachvili, M.B. Maple and H.C. Ku, J. Low Temp. Phys. {\bf 56} (1984) 601.
\bibitem{Malik_1983} S.K. Malik, R. Vijayaraghavan, W.E. Wallace and S.K. Dhar: J. Magn. Magn. Mater. {\bf 37} (1983) 303.
\bibitem{Galatanu_2002} A. Galatanu, E. Yamamoto, T. Okubo, M. Yamada, A Thamizhavel, T. Takeuchi, K. Sugiyama, Y. Inada and Y. Onuki: J. Phys.: Condens. Matter {\bf 15} (2003) S2187. 
\bibitem{Okubo_2003} T. Okubo, M. Yamada, A. Thamizhavel, S.K.Y. Inada, R. Settai, H. Harima, K. Takegahara, A. Galatanu, E. Yamamoto and Y. Onuki, J. Phys.: Condens. Matter: {\bf 15} (2003) L721.
\bibitem{Harima_2004} H. Harima and K. Takegahara, J. Magn. Magn. Mat. {\bf 272-276} (2004) 475.
\bibitem{Yamada_2004} M. Yamada, Y. Obiraki, T. Okubo, T. Shiromoto, Y. Kida, M. Shiimoto, H. Kohara, T. Yamamoto, D. Honda, A. Galatanu, Y. Haga, T. Takeuchi, K. Sugiyama, R. Settai, K. Kindo, S. K. Dhar, H. Harima and Y. \=Onuki: J. Phys. Soc. Japan {\bf 73} (2004) 2266.

\bibitem{Sampathkumaran_1985} E.V. Sampathkumaran, G. Kaindl, C. Laubschat, W. Krone and G. Wortmann: Phys. Rev. B {\bf 31} (1985) 3185.
\bibitem{Ku_1980} H. C. Ku, G. P. Meisner, F. Acker and D. C. Johnston: Solid State Commun. {\bf 35} (1990) 91.
\bibitem{Batista_2002} C.D. Batista, J. Bon{\v{c}}a and J.E. Gubernatis, Phys. Rev. Lett. {\bf 88} (2002) 187203.
\bibitem{Mielke_1993_2} A. Mielke and H. Tasaki: Commun. Math. Phys. {\bf 158} (1993) 31.
\bibitem{Tasaki1993} H. Tasaki: Phys. Rev. Lett. {\bf 73} (1993) 1158.
\bibitem{Penc_1996} K. Penc, H. Shiba, F. Mila and T. Tsukagoshi: Phys. Rev. B {\bf 54} (1996) 4056.
\bibitem{Guerrero_2001} M. Guerrero and R.M. Noack: Phys. Rev. B {\bf 63} (2001) 144423.
\bibitem{Yokoyama_1997} H. Yokoyama and S. Tokizaki: Physica B {\bf 230-232} (1997) 418.
\bibitem{Umrigar_1998} C.J. Umrigar, K.G. Wilson and J.W. Wilkins: Phys. Rev. Lett. {\bf 60} (1988) 1719.
\bibitem{Guerrero_1996}  M. Guerrero and R.M. Noack: Phys. Rev. B {\bf53} (1996) 3707.
\bibitem{Jo_1990} T. Jo and S. Imada: J. Phys. Soc. Jpn.  {\bf59} (1990) 2312.
\bibitem{Fujimori_1990} A. Fujimori, T. Takahashi, A. Okabe, M. Kasaya, and T. Kasuya: Phys. Rev. B {\bf41} (1990) 6783.
\bibitem{Yamaguchi_1995} K. Yamaguchi, H. Namatame, A. Fujimori, T. Koide, T. Shidara, M. Nakamura, A. Misu, H. Fukutani, M. Yuri, M. Kasaya, H. Suzuki and T. Kasuya: Phys. Rev. B {\bf 51} (1995) 13952.
\end{thebibliography}
\end{document}